%% file: main_v2.tex
\documentclass[prd,aps,twocolumn,preprintnumbers, showpacs, nofootinbib,superscriptaddress,notitlepage]{revtex4-1}
\usepackage{mathrsfs}
\usepackage{amsfonts}
\usepackage{amsmath}
\usepackage{slashed}
\usepackage{array}
\usepackage{verbatim}
\usepackage{epsfig}
\usepackage{graphicx}
\usepackage{color}
\usepackage[dvipsnames]{xcolor}
\usepackage{longtable}
\usepackage{multirow}

\usepackage[colorlinks,linkcolor=blue,citecolor=blue,urlcolor=blue]{hyperref}
\usepackage{csvsimple}
\usepackage{booktabs} 
\usepackage{siunitx}  
\usepackage{adjustbox} 
\usepackage{booktabs}
\usepackage{longtable}
\usepackage{multirow}
\renewcommand{\arraystretch}{1.2}

\definecolor{red}{rgb}{1,0,0}

\newcommand{\beq}{\begin{eqnarray}}
\newcommand{\eeq}{\end{eqnarray}}

\def\be{\begin{equation}}
\def\ee{\end{equation}}
\def\bea{\begin{eqnarray}}
\def\eea{\end{eqnarray}}

\begin{document}

\title{Unveiling the Strong Interaction origin of Baryon Masses with Lattice QCD
}

\collaboration{\bf{CLQCD Collaboration}}

\author{
\includegraphics[scale=0.30]{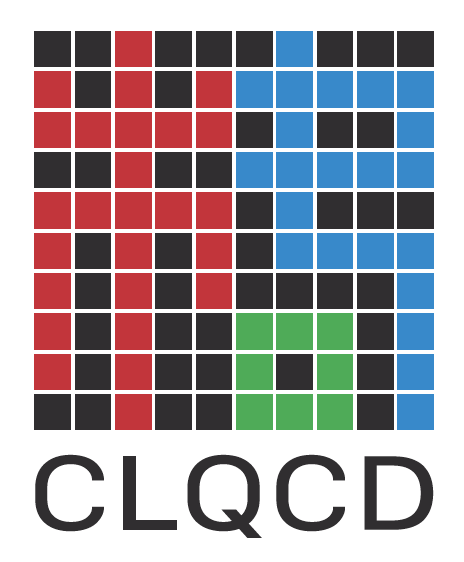}\\
Bolun Hu}
\affiliation{CAS Key Laboratory of Theoretical Physics, Institute of Theoretical Physics, Chinese Academy of Sciences, Beijing 100190, China}

\author{Haiyang Du}
\affiliation{University of Chinese Academy of Sciences, School of Physical Sciences, Beijing 100049, China}
\affiliation{CAS Key Laboratory of Theoretical Physics, Institute of Theoretical Physics, Chinese Academy of Sciences, Beijing 100190, China}

\author{Xiangyu Jiang}
\affiliation{CAS Key Laboratory of Theoretical Physics, Institute of Theoretical Physics, Chinese Academy of Sciences, Beijing 100190, China}

\author{Keh-Fei Liu}
\affiliation{University of Kentucky, Lexington, KY 40506, USA}
\affiliation{Nuclear Science Division, Lawrence Berkeley National Laboratory, Berkeley, 
California 94720, USA}

\author{Peng Sun}
\affiliation{Institute of Modern Physics, Chinese Academy of Sciences, Lanzhou, 730000, China}

\author{Yi-Bo Yang}
\email[Corresponding author: ]{ybyang@itp.ac.cn}
\affiliation{CAS Key Laboratory of Theoretical Physics, Institute of Theoretical Physics, Chinese Academy of Sciences, Beijing 100190, China}
\affiliation{University of Chinese Academy of Sciences, School of Physical Sciences, Beijing 100049, China}
\affiliation{School of Fundamental Physics and Mathematical Sciences, Hangzhou Institute for Advanced Study, UCAS, Hangzhou 310024, China}
\affiliation{International Centre for Theoretical Physics Asia-Pacific, Beijing/Hangzhou, China}

\date{\today}

\begin{abstract}
Both the Higgs mechanism and strong interactions contribute to the masses of visible matter, yet how the six Higgs-generated quark masses and uniform strong interaction strength determine the hundreds of hadron masses remains unclear. Additionally, the role of massless, flavor-neutral gluons on hadron mass formation is central to the unresolved Millennium Prize problem on the mass gap in Yang-Mills theory. Addressing these questions requires advanced simulations on state-of-the-art supercomputers using Lattice Quantum Chromodynamics (QCD), which offers a rigorous, non-perturbative definition of QCD solvable numerically~\cite{wilson:1974sk}. Here we present first-principles lattice QCD calculations using comprehensive gauge ensembles~\cite{CLQCD:2023sdb,CLQCD:2024yyn} that accurately predict ground state spin-1/2 and spin-3/2 baryon masses with light, strange, and charm quarks within 1\% of experimental values. At the \(\overline{\mathrm{MS}}\) 2 GeV scale, our results unveil two fundamental mass generation mechanisms for those baryon masses in QCD: 1) the flavor-dependent enhancement of Higgs contributions, 4-8 for light, 2-3 for strange, and 1.2-1.3 for charm quarks; and 2) the flavor-insensitive contribution 0.8-1.2 GeV from gluon quantum anomaly~\cite{Collins:1976yq,Nielsen:1977sy,Shifman:1978zn}. This breakthrough significantly advances our comprehension of strong interaction dynamics and the genesis of visible matter's mass.
\end{abstract}

\maketitle

Visible matter in the universe is composed of fundamental particles of the Standard Model (SM) of particle physics. During the first \(10^{-11}\) seconds after the Big Bang, all fundamental particles in SM are massless, leading to a state of nearly scale invariance in the universe. Following this period, the Higgs mechanism spontaneously breaks scale invariance and provides mass to these fundamental particles, resulting in the heaviest quark being approximately \(10^5\) times more massive than the lightest one. However, a second instance of spontaneous scale invariance breaking occurs at a relatively later stage likes $10^{-5}$ seconds in the universe, becoming an even more significant source of mass for visible matter, overshadowing the contribution of the Higgs mechanism.

This additional source of mass arises from the strong interaction between quarks and gluons, which is described by quantum chromodynamics (QCD) through the following Lagrangian:
\begin{align}
\mathcal{L}=-\frac{1}{2}\mathrm{Tr}G_{\mu}G^{\mu}+\sum_q \bar{q}\left(\gamma_{\mu}(\partial^{\mu}+igB^{\mu})+m_q\right)q,    
\end{align}
where \(G_{\mu\nu}=\partial_{\mu}B_{\nu}-\partial_{\nu}B_{\mu}+g[B_{\mu},B_{\nu}]\), with \(B_{\mu}\) representing a \(3\times 3\) Hermitian matrix field that interacts with itself via the QCD coupling constant \(g\). The matrices \(\gamma_{\mu}\) are the Dirac matrices, and different quark ``flavors" are depicted by distinct fermionic fields \(q\) with masses \(m_q\), which can be interpreted as the effective coupling between quarks and the Higgs boson.

The contribution from the Higgs mechanism to the mass of hadrons (such as protons and neutrons, which predominantly interact via strong forces) through a given quark flavor $q$ can be expressed as \( \sigma_{q,H}\equiv m_q \langle \bar{q} q \rangle_H\), where \(\langle \bar{q} q \rangle_H\) with $\langle O \rangle_H \equiv \langle H | O | H \rangle / \langle H | H \rangle$ can be understood as the condensate of quark and anti-quark within the hadron \(H\). 

Furthermore, quantum corrections in QCD cause both the gauge coupling \(g\) and the quark mass \(m_q\) to increase at longer distance scales, resulting in an additional breaking of scale invariance. Consequently, the energy-momentum tensor of matter, which couples matter to gravity, exhibits a non-vanishing trace, even as \(m_q\) approaches zero--this phenomenon is referred to as the ``trace anomaly". 

By combining sigma terms and the trace anomaly, one can decompose the total hadron mass as~\cite{Collins:1976yq,Nielsen:1977sy,Shifman:1978zn}, 
\begin{align}
m_H &= \sum_q \sigma_{q,H} +\gamma_m\sum_q \sigma_{q,H}+ \frac{\beta(\alpha_s)}{2\alpha_s} \langle G^{\mu\nu} G_{\mu\nu} \rangle_H,
\label{eqn:mass_decomp}  
\end{align}
where $m_H$ is the hadron mass, $\gamma_m = \frac{2}{\pi} \alpha_s + \mathcal{O}(\alpha_s^2)$ is the scale violation of quark mass anomalous dimension with $\alpha_s=\frac{g^2}{4\pi}$, and $\frac{\beta(\alpha_s)}{2\alpha_s}=(-\frac{11}{8\pi}+\frac{N_f}{12\pi})\alpha_s+\mathcal{O}(\alpha_s^2)$ is that of the effective strong coupling constant $\alpha_s$ with $N_f$ being the number of quark flavors. Thus, the quark trace anomaly term $\gamma_m$ enhances the quark sigma term contribution by an ${\cal O}(\alpha_s)$ correction, while the remaining part, associated with the gluon trace anomaly, is suppressed by an additional ${\cal O}(\alpha_s)$ factor in its contribution to the hadron mass via the double gluon exchange diagram. Additionally, there exist decompositions of the hadron rest energy, as discussed in Refs.~\cite{Ji:1994av,Yang:2018nqn,Lorce:2018egm,liu:2023cse}.

Since $\alpha_s$ is of order ${\cal O}(1)$ at the scale of the proton charge radius, the QCD interactions become strongly non-linear, potentially causing the power counting of $\alpha_s$ to break down. Ab initio calculations using a discretized Lagrangian on a hyper-cubic spacetime lattice, characterized by a finite lattice spacing \( a \) and a spatial size \( L \), transform QCD into a four-dimensional statistical physics system, lattice QCD~\cite{wilson:1974sk}. These calculations can precisely determine hadron masses and contributions from various sources through importance sampling~\cite{Duane:1987de}. The physical predictions are realized by taking the limits as \( a \rightarrow 0 \) and \( L \rightarrow \infty \).

\begin{widetext}

\begin{figure*}[h]
\includegraphics[width=\textwidth]{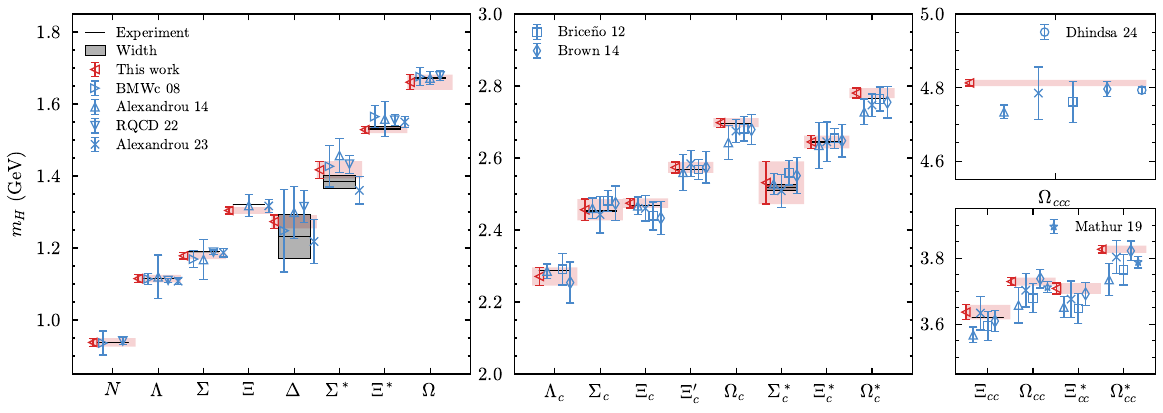}
\caption{\textbf{Comparison of Lattice QCD Baryon mass prediction and experimental values}. Red triangles and the accompanying red band represent our results, which are compared to previous studies (
        BMW 08 \cite{BMW:2008jgk}, 
        Alexandrou 14 \cite{Alexandrou:2014sha}, 
        RQCD 22
        \cite{RQCD:2022xux}
        Alexandrou 23 \cite{Alexandrou:2023dlu},         
        Briceno 12 \cite{Briceno:2012wt}, 
        Brown 14 \cite{Brown:2014ena}, 
        Mathur 19 \cite{Mathur:2018rwu}, 
        and Dhindsa 24 \cite{Dhindsa:2024erk}
        ) extrapolated to the continuum and physical quark masses. Experimental values are shown as black lines with shaded bands indicating decay widths. The figure consists of four panels, each corresponding to baryons containing 0 (left), 1 (middle), 2 (lower right), and 3 (upper right) charm quarks, respectively. In most cases, our predictions achieve the highest precision to date and exhibit good agreement with experimental values.}
    \label{fig:Hmass}
\end{figure*}
\end{widetext}

Based on the calculation with strange quark and two degenerated light quarks using  23,845 (for the proton and $\Delta$ baryon) and 17,725 (for the other baryons) measurements on the 3,340 configurations of 14 ensembles~\cite{CLQCD:2023sdb,CLQCD:2024yyn} in total, we predict the masses of $J^P=\frac{1}{2}^+$ and $\frac{3}{2}^+$ ground state baryons, and compare the results (red triangles and bands) with those in the literature~\cite{BMW:2008jgk,Alexandrou:2014sha,Briceno:2012wt,Brown:2014ena,Dhindsa:2024erk} (blue data points) and available experimental data (black lines with with shaded bands indicating decay widths to account for resonance effects in the measurements~\cite{BMW:2008jgk}), in Fig.~\ref{fig:Hmass}. There are also several lattice results without the continuum extrapolation which are not illustrated here~\cite{PACS-CS:2008bkb,Bietenholz:2011qq,Alexandrou:2017xwd,Liu:2009jc,PACS-CS:2013vie,Perez-Rubio:2015zqb,Li:2022vbc}. The black bands and blue data points are omitted where data are unavailable. This analysis includes baryons ranging from light to singly, doubly, and triply charmed species. In most cases, our predictions achieve the highest precision to date and exhibit good agreement with experimental values.

Accurate lattice QCD prediction requires good control on kinds of systematic uncertainties, as we address below:

1) Mismatch effects arising from the masses of the light and strange quarks: The ensembles~\cite{CLQCD:2023sdb,CLQCD:2024yyn} use different light quark masses ranging from 0.8 to 6.7 times the averaged mass of the up and down quarks. This allows us to interpolate the light quark mass to its physical value, using the SU($4|2$) partially quenched heavy baryon chiral perturbation theory ($\chi$PT) ansatz~\cite{Tiburzi:2005is,Yang:2018nqn} for the light baryons \( N \) and \( \Delta \), along with linear interpolation for the other baryons. Furthermore, we adjust the mass of the valence strange quark used to construct the baryon field to its physical value for each ensemble. By performing a joint fit across all the ensembles, we effectively eliminate the mismatch effects from the light and strange quarks loops.

2) Impact of tuning the charm quark mass and considering the missing charm quark loop: Similar to our approach with the strange quark, we adjust the valence charm quark mass to correspond with the physical mass of the \( D_s \) meson in each ensemble, and then eliminate the mismatch effects from the light and strange quark loops through a joint fit. As demonstrated in the newest review from Flavor lattice average group (FLAG)~\cite{flavourlatticeaveraginggroupflag:2024oxs}, which averages the lattice results from independent computations, the outcomes with and without the charm quark loop effects are consistent within the uncertainties. Notably, our prediction for \( m_{\Omega_{ccc}} \) aligns with that of Dhindsa 24~\cite{Dhindsa:2024erk}, which included the charm quark loop in their calculations using different discretizations and also extrapolated to the continuum limit as \( a \rightarrow 0 \), similar to our calculation here.

3) Continuum and infinite volume extrapolation: We find that the linear \( a^2 \) correction adequately describes the masses of baryons without the charm quark across different lattice spacings \( a \). Additionally, the \( a^4 \) correction is crucial for capturing the \( a \) dependence of certain charmed baryons. Consequently, we retain the \( a^4 \) term in the continuum extrapolation for all charmed baryons, which results in our predictions exhibiting relatively larger uncertainties compared to some cases without valence charm quark. We also include the leading-order finite volume correction in our fit, and its impact is shown to be smaller than the statistical uncertainties.

4) Isospin symmetry breaking (ISB) and quantum electrodynamics (QED) effects: Naive power counting indicates that both effects are at the 1\% level and then unlikely to alter the conclusions presented in this work. Therefore, we defer their investigation to future studies.

More evidence validating above procedure are provided in the supplementary information.

\begin{figure*}[t]
    \includegraphics[width=0.98\textwidth]{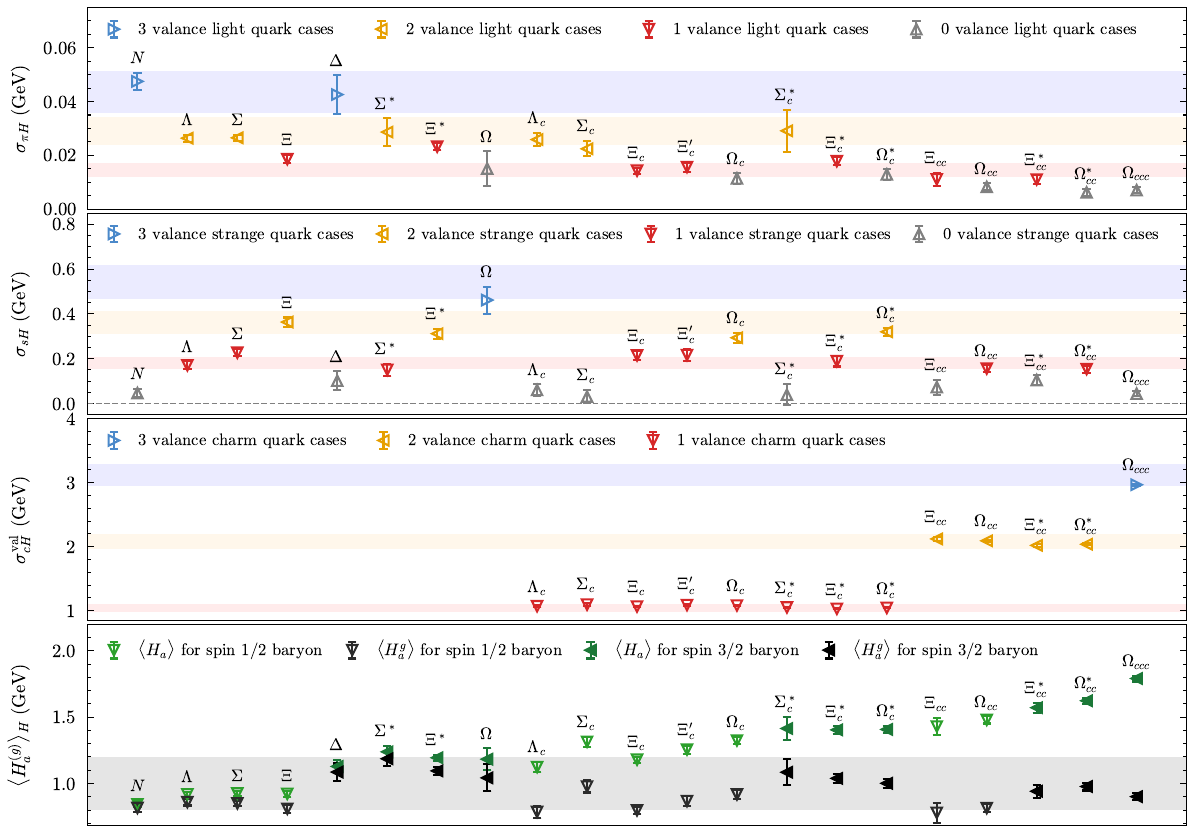}
    \caption{\textbf{Quark sigma terms and trace anomaly contributions to baryon masses}. The top, middle-upper and middle-lower panels show the $\sigma$ terms for light, strange, and charm quarks, respectively, with colors indicating the number of valence quarks: gray (0), red (1), yellow (2), and blue (3). Most colored data points align with their corresponding colored bands, following the scaling law based on quark numbers. The bottom panel presents the trace anomaly $\langle H_a \rangle_H$ (green triangles) and the gluon trace anomaly $\langle H^g_a \rangle_H$ (black ones), with hollow triangles for spin-$1/2$ baryons and filled triangles for spin-$3/2$ baryons. A gray band indicates the range of $\langle H^g_a \rangle_H$ values for all displayed baryons.}
    \label{fig:Hm_4in1}
\end{figure*}

The scalar matrix elements (ME) $\langle \bar{q}q\rangle_H$ can be obtained from the quark mass dependence of the hadron mass via the Feynman-Hellman theorem~\cite{RQCD:2022xux,Agadjanov:2023efe,Borsanyi:2020bpd,Copeland:2021qni,Frink:2004ic,Alexandrou:2014sha,Gupta:2021ahb,MartinCamalich:2010fp,PACS-CS:2009cvn,Ren:2014vea,Ren:2012aj,Semke:2012gs,Yang:2018nqn,Young:2009ps,Alvarez-Ruso:2013fza,BMW:2011sbi,QCDSF-UKQCD:2011qop,An:2014aea,Junnarkar:2013ac,Takeda:2010cw,Toussaint:2009pz}, or direct calculation of the scalar operator $\bar{q}q$ ME in the hadron state $H$~\cite{Alexandrou:2019brg,Durr:2015dna,Yang:2015uis,Bali:2016lvx,XQCD:2013odc,Abdel-Rehim:2016won,QCDSF:2011mup}. Since the consistency of the two strategies has been validated for both light and charm quarks in the CLQCD ensembles after performing the continuum extrapolation~\cite{CLQCD:2023sdb, CLQCD:2024yyn}, we opt for the Feynman-Hellman theorem, which avoids the complications associated with the direct ME calculation. The consistency of the light quark matrix element in the octet light baryon using two approaches is also verified using different fermion discretizations~\cite{Yang:2018nqn,Petrak:2023qhx}.

The combined contribution $\sigma_H\equiv \sum_q \sigma_{q,H}$ of all the quark flavors in the nucleon case corresponds to the effective Higgs-nucleon couplings \cite{Hisano:2010ct, Hisano:2011cs}, and is crucial for constraining dark matter-nucleon interactions in direct detection experiments \cite{Ellis:2008hf, Crivellin:2013ipa}. After $\sigma_H$ is obtained, the trace anomaly contribution and its glue part can be extracted from Eq.~(\ref{eqn:mass_decomp}),
\begin{align}
    \langle H_{a} \rangle_H &= m^{\rm phys}_H - \sigma_H, \nonumber\\
    \langle H^g_{a} \rangle_H &= \langle H_{a} \rangle_H - \gamma_m \sigma_H=m^{\rm phys}_H - (1+\gamma_m)\sigma_H.\label{eqn:Ha}
\end{align}

1) Light quark sigma term $\sigma_{\pi H}\equiv \sigma_{u H}+\sigma_{d H}$: The upper left panel of Fig.~\ref{fig:Hm_4in1} displays the light quark $\sigma$ terms, representing the direct contributions of light quark masses to the baryon mass. Some of the enhancement originates solely from the light quark loop (a gluon splits into a quark and anti-quark pair which combines into another gluon at a later time), as observed in the cases of $\Omega^{(*)}_{(c/cc/ccc)}$ baryons (gray triangles), which do not have any valence light quark. However, the $\sigma_{\pi H}$ of other baryons can be significantly larger, as $\langle \bar{q}q\rangle_H$ of the valence quark can be about four times greater than the number of valence light quarks, using the average up and down quark mass of 3.39(4) MeV at $\overline{\mathrm{MS}}$ 2 GeV~\cite{flavourlatticeaveraginggroupflag:2024oxs,Bruno:2019vup, RBC:2014ntl, BMW:2010ucx, BMW:2010skj, McNeile:2010ji, Bazavov:2010yq}. This additional enhancement is likely associated with the connected-sea mechanism~\cite{Liu:2012ch}, where a gluon can split into a quark that moves forward in time and an antiquark that combines with a quark from the past to form another gluon, thereby introducing a surplus of quarks and antiquarks. In terms of the Feynman diagram, this mechanism is akin to a scenario where a valence light quark travels back in time and returns due to interactions with the gluon.


2) Strange quark sigma term $\sigma_{s H}$: When the quark mass reaches the magnitude of the strange quark mass $92.4(1.0)$ MeV at $\overline{\mathrm{MS}}$ 2 GeV~\cite{flavourlatticeaveraginggroupflag:2024oxs,Bruno:2019vup, RBC:2014ntl, BMW:2010ucx, BMW:2010skj, McNeile:2010ji, MILC:2009ltw}, the enhancements from both the quark loop and connected-sea mechanisms diminish, as the gluon must provide significantly more energy for the related processes to occur. Consequently, as observed in the upper-right panel of Fig.~\ref{fig:Hm_4in1}, the enhancement factor decreases to approximately 1.5, regardless of the number of valence strange quarks, angular momentum, or the flavor of the remaining quarks. This factor is considerably smaller compared to that of the light quark, indicating that the strange quark exhibits less relativistic behavior.


3) Charm quark sigma term $\sigma_{c H}$: The lower-left panel of Fig.~\ref{fig:Hm_4in1} illustrates that the charm quark scenario is notably more consistent, with the valence $\sigma_{c H}$ being directly proportional to the charm quark by a factor of approximately 1. In comparison to the charm quark mass of 1.093(7) GeV at $\overline{\mathrm{MS}}$ 2 GeV~\cite{flavourlatticeaveraginggroupflag:2024oxs, McNeile:2010ji,Bussone:2023kag, Yang:2014sea, Nakayama:2016atf, Petreczky:2019ozv, Heitger:2021apz}, the ratio may be slightly smaller than the number of valence charm quark masses. Since the quark mass decreases at higher energy scales, this suggests that an inherent scale—required to ensure that $\langle\bar{c}c\rangle$ matches the number of valence charm quarks—must be higher as the number of valence charm quarks increases. This scale ultimately reaches 2.9 GeV for the heaviest $\Omega_{ccc}$, which contains three valence charm quarks.

The charm sea quark contribution to the baryon mass is neglected here, not only because our gauge ensemble includes only light and strange quarks, but also due to the heavy quark relation $m_Q\bar{Q}Q=-\frac{\alpha_s}{12\pi}F^2$~\cite{Shifman:1978zn,Tarrach:1981bi}, which implies that the heavy sea quark contribution is canceled by the \(N_f\) dependence of the gluon trace anomaly at leading order, leading to the decoupling theorem~\cite{Appelquist:1974tg}. One can also calculate this contribution based on the trace anomaly, as discussed below.
 

4) Trace anomaly contribution: By subtracting the total $\sigma$ term from the baryon mass using Eq.~\eqref{eqn:Ha}, the values of $\langle H_a \rangle_H$ (represented by hollow triangles) for different hadrons $H$ are displayed in the lower-right panel of Fig.~\ref{fig:Hm_4in1}. It is evident that $\langle H_a \rangle_H$ increases with the number of heavy valence quarks, resulting in a twofold enhancement from the lightest baryon $\langle H_a \rangle_N$ to the heaviest one $\langle H_a \rangle_{\Omega_{ccc}}$ that we have examined. However, what is particularly noteworthy is that $\langle H^g_a \rangle_H$ (depicted by filled triangles) becomes less dependent on the valence flavor composition and converges around 1 GeV when we further subtract the anomalous sigma term $\gamma_m\sigma_H$ with $\gamma_m=0.295$ at $\overline{\mathrm{MS}}$ 2 GeV~\cite{particledatagroup:2022pth}. Interestingly, the spin-$3/2$ baryons exhibit slightly higher gluon trace anomaly contributions compared to their spin-$1/2$ counterparts.

Based on the relation $m_Q\bar{Q}Q=-\frac{\alpha_s}{12\pi}F^2$, one can estimate the heavy sea quark contribution to all kinds of the baryon we studied here to be 74(15) MeV at the leading order of the strong coupling and consistent with the direct calculations using gauge ensembles including a dynamical heavy quark~\cite{Borsanyi:2020bpd}.

More details of the $\sigma_{q,H}$ calculation can also be found in the supplemental information.

In summary, the ground state masses of $\frac{1}{2}^+$ and $\frac{3}{2}^+$ baryons with the lightest four flavors are predicted by effectively controlling all systematic uncertainties, except for the ISB and QED effects, along with the contributions from Higgs and gluon couplings of the quarks. 
Despite the Higgs coupling to the charm quark is hundreds of times that of the light quark, its impact on visible matter is largely disregarded when the valence heavy quark in the baryon decays to the light quark through weak interactions. In contrast, the baryon mass resulting from the gluon trace anomaly persists. Consequently, the gluon trace anomaly, which is insensitive to hadrons, along with the heavy quark loop contribution which is proportional to the gluon trace anomaly one and insensitive to the Higgs coupling, emerges as the primary source of visible matter. Furthermore, the strong interaction also mitigates the Higgs coupling to the baryon through various quark flavors, exhibiting a more substantial enhancement for lighter quarks.

Recent direct calculations of the gluon trace anomaly contribution~\cite{He:2021bof} reveal that this contribution is significantly smaller for the pion—QCD’s pseudo-Goldstone boson—and vanishes in the limit of negligible quark masses. Complementary studies of the pion’s spatial distribution~\cite{He:2021bof} and form factors~\cite{Wang:2024lrm,Cao:2024zlf} further suggest the presence of a negative mass ``hollow" at its center, arising from the trace anomaly. This distinctive feature differentiates the pion’s mass structure from that of other hadrons. Insights into these phenomena can be indirectly derived through the off-diagonal components of the energy-momentum tensor (EMT) form factors~\cite{Hackett:2023rif,Hackett:2023nkr}, guided by EMT sum rules~\cite{ji:2021mtz,liu:2023cse}, and may also be probed experimentally via near-threshold $J/\Psi$ photoproduction~\cite{Kharzeev:1995ij,Hatta:2018ina,Duran:2022xag}.
These findings highlight the need for deeper exploration to elucidate the unique mechanisms by which strong interactions govern the origin of visible mass in hadronic systems.

\section*{Acknowledgment}
We thank the CLQCD collaborations for providing us their gauge configurations with dynamical fermions~\cite{CLQCD:2023sdb}, which are generated on HPC Cluster of ITP-CAS, the Southern Nuclear Science Computing Center(SNSC) and the Siyuan-1 cluster supported by the Center for High Performance Computing at Shanghai Jiao Tong University. 
We thank Hanyang Xing for providing two-point functions using distillation method for comparison, and Gunnar Bali, Ying Chen, Hengtong Ding, Xu Feng, Jianxiao Gong, Chuan Liu, Liuming Liu, Zhaofeng Liu, Ji-Hao Wang, Kuan Zhang, and the other CLQCD members for valuable comments and suggestions.
The calculations were performed using PyQUDA~\cite{Jiang:2024lto} and Chroma ~\cite{Edwards:2004sx} with QUDA~\cite{Clark:2009wm,Babich:2011np,Clark:2016rdz}, through HIP programming model~\cite{Bi:2020wpt}. The numerical calculation were carried out on the ORISE Supercomputer, HPC Cluster of ITP-CAS and Advanced Computing East China Sub-center. This work is supported in part by National Key R\&D Program of China No.2024YFE0109800, NSFC grants No. 12293060, 12293062, 12435002, 12293065 and 12047503, the Strategic Priority Research Program of Chinese Academy of Sciences, Grant No.\ XDB34030303 and YSBR-101, and also the science and education integration young faculty project of University of Chinese Academy of Sciences.

\bibliography{ref}

\clearpage

\include{sm}

\end{document}

%% file: sm.tex
\appendix
\begin{widetext}

\section*{Supplemental materials}

\begin{table*}[t]
  \centering
  \caption{ Lattice size $\tilde{L}^3\times \tilde{T}$, gauge coupling $\hat{\beta}$, bare quark mass parameters $\tilde{m}^{\rm b}_{l,s}$, the corresponding pseudoscalar mass $m_{\pi, \eta_s}$, and the statistics information.}\label{tab:ensembles} 
    \begin{tabular}{l | c c c c c c c c c c c}
      \hline\hline
       Symbol& $\tilde{L}^3\times \tilde{T}$ & $\hat{\beta}$ & $a$ (fm) & $\tilde{m}^{ b}_l$ & $\tilde{m}^{ b}_s$ & $m_{\pi}$(MeV) & $m_{\eta_s}$(MeV) & $m_{\pi}L$ & $n_{\rm cfg}$ & $n_{\rm src}^l$ & $n_{\rm src}^{s,c}$ \\
      \hline
      C24P34 & $24^3\times 64$ & 6.20 & 0.10521(11)(62) & -0.2770 & -0.2310 & 340.2(1.7) & 748.61(75) & 4.38 & 199 & 32 & 32 \\
C24P29 & $24^3\times 72$ &  & & -0.2770 & -0.2400 & 292.3(1.0) & 657.83(64) & 3.75 & 760 & 3 & 3 \\
C32P29 & $32^3\times 64$ & & & -0.2770 & -0.2400 & 293.1(0.8) & 658.80(43) & 5.01 & 489 & 3 & 3 \\
C32P23 & $32^3\times 64$ & & & -0.2790 & -0.2400 & 227.9(1.2) & 643.93(45) & 3.91 & 400 & 3 & 3 \\
C48P23 & $48^3\times 96$ & & & -0.2790 & -0.2400 & 224.1(1.2) & 644.08(62) & 5.79 & 60 & 3 & 3 \\
C48P14 & $48^3\times 96$ & & & -0.2825 & -0.2310 & 136.4(1.7) & 706.55(39) & 3.56 & 136 & 48 & 3 \\
\hline
E28P35 & $28^3\times 64$ & 6.308 & 0.08970(26)(53) & -0.2490 & -0.2170 & 351.4(1.4) & 717.94(93) & 4.43 & 142 & 4 & 4 \\
\hline
F32P30 & $32^3\times 96$ & 6.41 & 0.07751(14)(45) & -0.2295 & -0.2050 & 300.4(1.2) & 675.98(97) & 3.81 & 250 & 3 & 3 \\
F48P30 & $48^3\times 96$ & & & -0.2295 & -0.2050 & 302.7(0.9) & 674.76(58) & 5.72 & 99 & 3 & 3 \\
F32P21 & $32^3\times 64$ & & & -0.2320 & -0.2050 & 210.3(2.3) & 658.79(94) & 2.67 & 194 & 3 & 3 \\
F48P21 & $48^3\times 96$ & & & -0.2320 & -0.2050 & 207.5(1.1) & 661.94(64) & 3.91 & 75 & 3 & 3 \\
F64P12 & $64^3\times 128$ & & & -0.2336 & -0.2030 & 122.8(0.9) & 679.90(30) & 3.09 & 109 & 4 & 4 \\
\hline
G36P29 & $36^3\times 108$ & 6.498 & 0.06884(18)(41) & -0.2150 & -0.1926 & 297.2(0.9) & 693.05(46) & 3.68 & 270 & 4 & 4 \\
\hline
H48P32 & $48^3\times 144$ & 6.72 & 0.05198(20)(31) & -0.1850 & -0.1700 & 316.6(1.0) & 691.88(65) & 4.06 & 157 & 12 & 12 \\
      \hline\hline
    \end{tabular}
\end{table*}

\subsection{Details of the ensembles}
The ensembles used in this work are summarized in Table~\ref{tab:ensembles}. They were generated using a tadpole-improved tree-level Symanzik (TITLS) gauge action with a 2+1 flavor tadpole-improved tree-level Clover (TITLC) fermion action.

The TITLS gauge action, denoted as \( S_g \), is defined as
\[
S_g = \frac{1}{N_c} \mathrm{Re} \sum_{x, \mu < \nu} \mathrm{Tr} \left[ 1 - \hat{\beta} \left( \mathcal{P}^U_{\mu, \nu}(x) + \frac{c_1 \mathcal{R}^U_{\mu, \nu}(x)}{1 - 8c_1^0} \right) \right] \;,
\]
where \( N_c = 3 \), 
\[
\mathcal{P}^{U}_{\mu, \nu}(x) = U_\mu(x) U_\nu(x + a \hat{\mu}) U^{\dagger}_\mu(x + a \hat{\nu}) U^{\dagger}_\nu(x)
\]
and
\[
\mathcal{R}^{U}_{\mu, \nu}(x) = U_\mu(x) U_\mu(x + a \hat{\mu}) U_\nu(x + 2a \hat{\mu}) U^{\dagger}_\mu(x + a \hat{\mu} + a \hat{\nu}) U^{\dagger}_\mu(x + a \hat{\nu}) U^{\dagger}_\nu(x)\;,
\]
with the link variable \( U_\mu(x) \) given by
\[
U_{\mu}(x) = P \left[ \exp \left( i g_0 \int_{x + \hat{\mu}a}^x \mathrm{d}y B_{\mu}(y) \right) \right],
\]
and
\(
\hat{\beta} = (1 - 8c_1^0) \frac{6}{g_0^2 u_0^4} \equiv \frac{10}{g_0^2 u_0^4},
\)
with \( c_1^0 = -\frac{1}{12} \), \( c_1 = \frac{c_1^0}{ u_0^2} \), and
\[
u_0 = \left\langle \frac{\mathrm{Re} \mathrm{Tr} \sum_{x, \mu < \nu} \mathcal{P}^{U}_{\mu \nu}(x)}{6 N_c \tilde{V}} \right\rangle^{1/4}
\]
being the tadpole improvement factor. Here, \( \tilde{V} = \tilde{L}^3 \times \tilde{T} \) is the dimensionless 4-D volume of the lattice, and we use \( \tilde{O} \) for the dimensionless value of any quantity \( O \).

The TITLC fermion action employs a 1-step stout smeared link \( V \) with smearing parameter \( \rho = 0.125 \), and is given by
\[
S_q(m) = \sum_{x, \mu = 1, \dots, 4, \eta = \pm} \bar{\psi}(x + \eta \hat{\mu} a) \frac{1 - \eta \gamma_{\mu}}{2} V_{\eta \mu}(x) \psi(x) + \sum_x \bar{\psi}(x) \left[ -(4 + ma) \delta_{y,x} + c_{\rm sw} \sigma^{\mu \nu} g_0 F^V_{\mu \nu} \right] \psi(x),
\]
where \( c_{\rm sw} = \frac{1}{v_0^3} \) with \( v_0 = \left\langle \frac{\mathrm{Re} \mathrm{Tr} \sum_{x, \mu < \nu} \mathcal{P}^{V}_{\mu \nu}(x)}{6 N_c \tilde{V}} \right\rangle^{1/4} \), and
\[
F^V_{\mu \nu} = \frac{i}{8 a^2 g_0} \left( \mathcal{P}^V_{\mu, \nu} - \mathcal{P}^V_{\nu, \mu} + \mathcal{P}^V_{\nu, -\mu} - \mathcal{P}^V_{-\mu, \nu} \right.
+ \mathcal{P}^V_{-\mu, -\nu} - \mathcal{P}^V_{-\nu, -\mu} + \mathcal{P}^V_{-\nu, \mu} - \mathcal{P}^V_{\mu, -\nu}).
\]
Based on the joint fit on subsets of these ensembles with several lattice spacing, quark masses and volumes~\cite{CLQCD:2023sdb,CLQCD:2024yyn}, the physical up, down, strange and charm quark masses and also related CKM matrix elements $V_{u(c)d(s)}$ agree with the current lattice averages well~\cite{flavourlatticeaveraginggroupflag:2024oxs}. These ensembles have been also used to study the hadron spectrum~\cite{Xing:2022ijm,Liu:2022gxf,Liu:2023feb,Liu:2023pwr,Yan:2024yuq} and structures~\cite{Zhang:2021oja,Han:2024min,Meng:2024gpd,Meng:2024nyo}.

\subsection{Interpolation fields, 2-point functions and ground state baryon mass extraction\label{sec:interpolation fields}}
For the interpolation fields of the baryons, we use $\epsilon_{abc} P^+\, \left[ (q^1_a)^T C \gamma^5 q^2_b \right]q^3_c$ and $\epsilon_{abc} P^+\, \left[ (q^1_a)^T C \gamma^{\mu} q^2_b \right]q^3_c$ for the spin-1/2 and 3/2 particles, respectively. Note that some of the particles require linear combination of different permutations of three quarks. The interpolation fields for the baryons we investigated in this work are summarized in Table~\ref{tab:interpolation_fields}. For baryons with light (up and/or down) quarks, we maximize the number of up quarks in the interpolation field, as the other configurations are equivalent due to isospin symmetry.

As presented in Table~\ref{tab:ensembles}, we computed the Coulomb gauge fixed source propagators at $n_{\rm src}^l$ time slices for each configuration at three light quark masses including the unitary one, and at $n_{\rm src}^{s,c}$ for three strange and charm quark masses. We have 23,845 (for the proton and $\Delta$ baryon) and 17,725 (for the other baryons) measurements on the 3,340 configurations of 14 ensembles in total.

\begin{table*}[t]
\centering

\renewcommand{\arraystretch}{1.2}
\begin{tabular}{c c c c c c}
\hline\hline
Baryon & Quark Content & Interpolating Field & \( I \) & \( I_z \) & $J$ \\
\hline
\( p \) & \( uud \) & \( \epsilon_{abc}P^+(d_a^T C \gamma^5 u_b) u_c \) & \( \tfrac{1}{2} \) & \( +\tfrac{1}{2} \) & \multirow{11}{*}{$\frac{1}{2}$} \\
\( \Lambda \) & \( uds \) & \(  \epsilon_{abc}P^+\left[2(u_a^T C \gamma^5 d_b) s_c + (u_a^T C \gamma^5 s_b) d_c - (d_a^T C \gamma^5 s_b) u_c\right] \) & \( 0 \) & \( 0 \) & \\
\( \Sigma^{+} \) & \( uus \) & \( \epsilon_{abc}P^+(s_a^T C \gamma^5 u_b) u_c \) & \( 1 \) & \( +1 \) & \\
\( \Xi^{0} \) & \( uss \) & \( \epsilon_{abc}P^+(u_a^T C \gamma^5 s_b) s_c \) & \( \tfrac{1}{2} \) & \( +\tfrac{1}{2} \) & \\
\( \Lambda_c \) & \( udc \) & \(  \epsilon_{abc}P^+\left[2(u_a^T C \gamma^5 d_b) c_c + (u_a^T C \gamma^5 c_b) d_c - (d_a^T C \gamma^5 c_b) u_c\right] \) & \( 0 \) & \( 0 \) & \\
\( \Sigma_c^{++} \) & \( uuc \) & \( \epsilon_{abc}P^+(c_a^T C \gamma^5 u_b) u_c \) & \( 1 \) & \( +1 \) & \\
\( \Xi_c^{+} \) & \( usc \) &\(  \epsilon_{abc}P^+\left[2(u_a^T C \gamma^5 s_b) c_c + (u_a^T C \gamma^5 c_b) s_c - (s_a^T C \gamma^5 c_b) u_c\right] \)  & \( \tfrac{1}{2} \) & \( +\tfrac{1}{2} \) & \\
\( \Xi_c^{\prime +} \) & \( usc \) & \(  \epsilon_{abc}P^+\left[(u_a^T C \gamma^5 c_b) s_c + (s_a^T C \gamma^5 c_b) u_c\right] \) & \( \tfrac{1}{2} \) & \( +\tfrac{1}{2} \) & \\
\( \Omega_c^{0} \) & \( ssc \) & \( \epsilon_{abc}P^+(c_a^T C \gamma^5 s_b) s_c \) & \( 0 \) & \( 0 \) & \\
\( \Xi_{cc}^{++} \) & \( ucc \) & \( \epsilon_{abc}P^+(u_a^T C \gamma^5 c_b) c_c \) & \( \tfrac{1}{2} \) & \( +\tfrac{1}{2} \) & \\
\( \Omega_{cc}^{+} \) & \( scc \) & \( \epsilon_{abc}P^+(s_a^T C \gamma^5 c_b) c_c \) & \( 0 \) & \( 0 \) & \\
 \hline
\( \Delta^{++} \) & \( uuu \) & \( \epsilon_{abc}P^+(u_a^T C \gamma^\mu u_b) u_c \) & \( \tfrac{3}{2} \) & \( +\tfrac{3}{2} \) & \multirow{10}{*}{$\tfrac{3}{2}$} \\
\( \Sigma^{*+} \) & \( uus \) & \(  \epsilon_{abc}P^+\left[(u_a^T C \gamma^\mu u_b) s_c + 2(s_a^T C \gamma^\mu u_b) u_c\right] \) & \( 1 \) & \( +1 \) & \\
\( \Xi^{*0} \) & \( uss \) & \(  \epsilon_{abc}P^+\left[2(s_a^T C \gamma^\mu u_b) s_c + (s_a^T C \gamma^\mu s_b) u_c\right] \) & \( \tfrac{1}{2} \) & \( +\tfrac{1}{2} \) & \\
\( \Omega^{-} \) & \( sss \) & \( \epsilon_{abc}P^+(s_a^T C \gamma^\mu s_b) s_c \) & \( 0 \) & \( 0 \) & \\
\( \Xi_c^{*+} \) & \( usc \) & \(  \epsilon_{abc}P^+\left[(u_a^T C \gamma^\mu s_b) c_c + (s_a^T C \gamma^\mu c_b) u_c + (c_a^T C \gamma^\mu u_b) s_c\right] \) & \( \tfrac{1}{2} \) & \( +\tfrac{1}{2} \) & \\
\( \Omega_c^{*0} \) & \( ssc \) & \(  \epsilon_{abc}P^+\left[2(s_a^T C \gamma^\mu c_b) s_c + (s_a^T C \gamma^\mu s_b) c_c\right] \) & \( 0 \) & \( 0 \) & \\
\( \Sigma^{*++}_c \) & \( uuc \) & \(  \epsilon_{abc}P^+\left[(u_a^T C \gamma^\mu u_b) c_c + 2(c_a^T C \gamma^\mu u_b) u_c\right] \) & \( 1 \) & \( +1 \) & \\
\( \Xi_{cc}^{*++} \) & \( ucc \) & \(  \epsilon_{abc}P^+\left[2(c_a^T C \gamma^\mu u_b) c_c + (c_a^T C \gamma^\mu c_b) u_c\right] \) & \( \tfrac{1}{2} \) & \( +\tfrac{1}{2} \) & \\
\( \Omega_{cc}^{*+} \) & \( scc \) & \(  \epsilon_{abc}P^+\left[2(c_a^T C \gamma^\mu s_b) c_c + (c_a^T C \gamma^\mu c_b) s_c\right] \) & \( 0 \) & \( 0 \) & \\
\( \Omega_{ccc}^{++} \) & \( ccc \) & \( \epsilon_{abc}P^+(c_a^T C \gamma^\mu c_b) c_c \) & \( 0 \) & \( 0 \) & \\
\hline\hline
\end{tabular}
\caption{Interpolating fields for baryons.}
\label{tab:interpolation_fields}
\end{table*}

 Taking the nucleon as example, the the wall-to-point two-point functions we used in the this work is, explicitly recovering the coordinate and Dirac indices,
\begin{multline}
C_{2,N}^{wp}(y^0,x^0) = \sum_{\vec{\boldsymbol{x}},\vec{\boldsymbol{y}}} \left\langle \chi_{N}^\alpha(y) \bar{\chi}_{N}^\alpha(x) \right\rangle 
= \sum_{\vec{\boldsymbol{y}}} \epsilon_{abc} \epsilon_{a'b'c'} 
\Bigg\langle\left( C \gamma_5 \right)^{\alpha' \beta'} \left( C \gamma_5 \right)^{\alpha \beta} 
\left( P^{\pm} \right)^{\gamma \gamma'} S^C_d(y,x^0)_{\substack{\alpha' \alpha \\ a' a}} \\
\times \left[ S^C_u(y,x^0)_{\substack{\beta' \beta \\ b' b}} S^C_u(y,x^0)_{\substack{\gamma' \gamma \\ c' c}} 
- S^C_u(y,x^0)_{\substack{\beta' \gamma \\ b' c}} S^C_u(y,x^0)_{\substack{\gamma' \beta \\ c' b}} \right]\Bigg\rangle\;,
\end{multline}
where the Coulomb gauge-fixed wall propagator is defined by $S_q^C\left(y , x^0 \right) \equiv \sum_{\vec{\boldsymbol{x}}} S\left(y, x ;U_C;\tilde{m}_q^b\right)$ as used in Ref.~\cite{CLQCD:2023sdb}, and $S\left(y, x;U_C\right) = \psi\left(y;U_C\right) \bar{\psi}\left(x;U_C\right)$
is the standard quark propagator computed with the gauge configuration \(U\). The Coulomb gauge-fixed configuration \(U_C\) satisfies the discretized gauge-fixing condition $\operatorname{Im}\left[\sum_{i=1}^3\left(U_C(x) - U_C\left(x - a \hat{n}_i\right)\right)\right] = 0$.

\begin{figure*}[t]
\centering
   \begin{center}
      \begin{tabular}{@{}cccc@{}}
     \includegraphics[width=.47\textwidth]{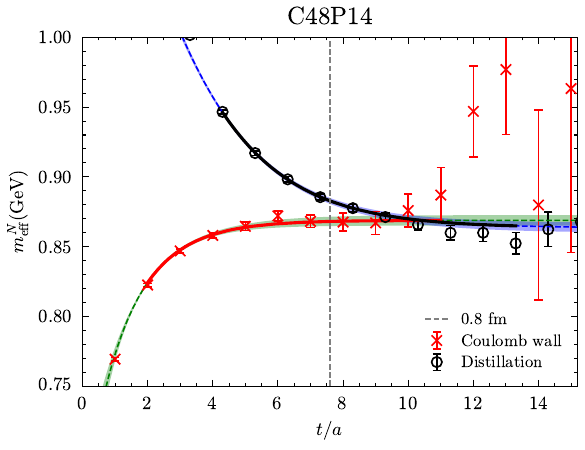} 
     \includegraphics[width=.47\textwidth]{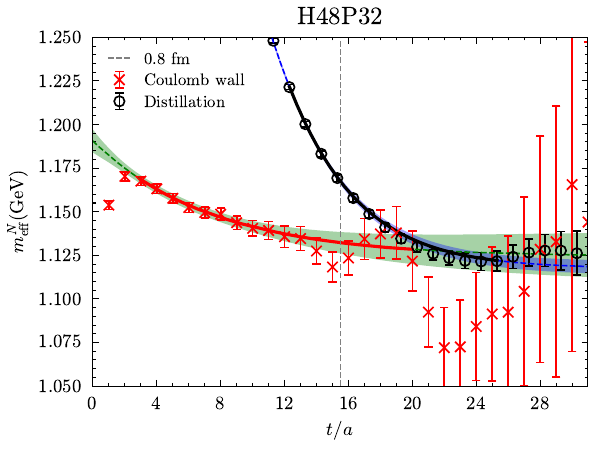} 
  \end{tabular}
  \caption{The effective mass obtained from the two-point correlation functions (2pt) of the nucleon in two representative ensembles: the physical point ensemble C48P14 (left panel) and the ensemble with the finest lattice spacing, H48P32 (right panel). The effective mass derived from the 2pt correlation functions (using a Coulomb wall source and point sink, indicated by red crosses) is compared with the results obtained using the distillation method (shown as black circles). Both methods exhibit good agreement in the plateau region, despite different initial trends with respect to \(t/a\). The green bands represent the ground state mass that we extracted.}
   \label{fig:2pt_fit_1}
   \end{center}
\end{figure*}

\begin{figure*}[t]
\centering
   \begin{center}
      \begin{tabular}{@{}cccc@{}}
     \includegraphics[width=.47\textwidth]{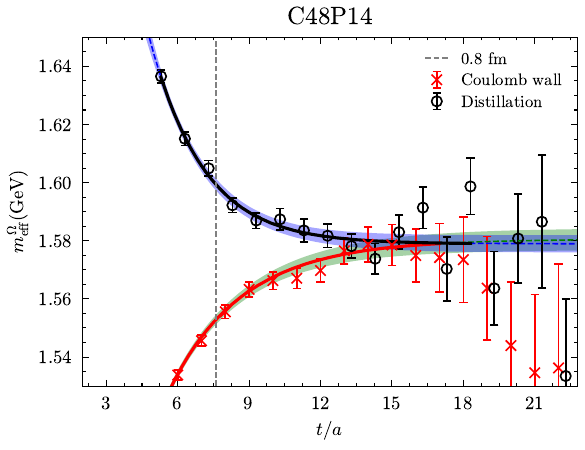} 
     \includegraphics[width=.47\textwidth]{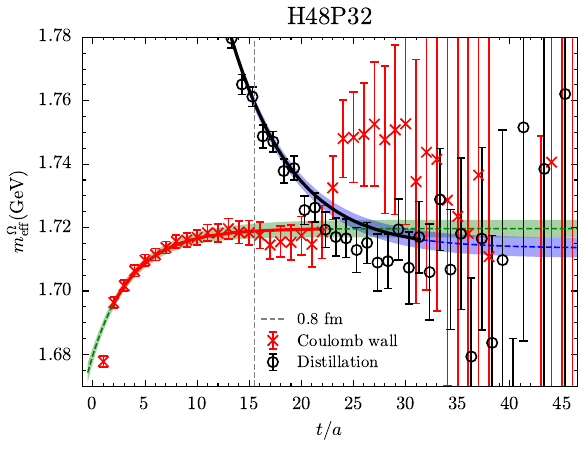} \\
   \end{tabular}
  \caption{Similar figures for the $\Omega$ baryon.}
   \label{fig:2pt_fit_2}
   \end{center}
\end{figure*}

\begin{figure*}[t]
\centering
   \begin{center}
      \begin{tabular}{@{}cccc@{}}
     \includegraphics[width=.47\textwidth]{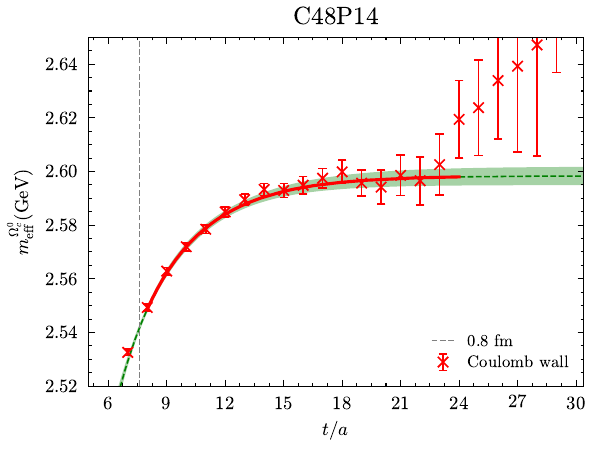} 
     \includegraphics[width=.47\textwidth]{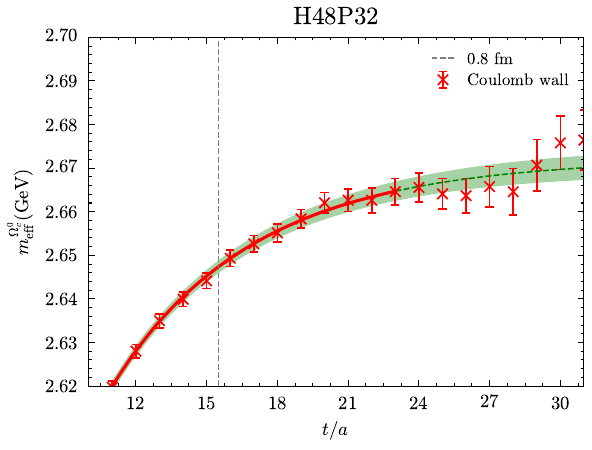} \\
   \end{tabular}
  \caption{Similar figures for the $\Omega_c$ baryon, while the distillation case is not available.}
   \label{fig:2pt_fit_3}
   \end{center}
\end{figure*}

In Fig.~\ref{fig:2pt_fit_1}-\ref{fig:2pt_fit_3}, we present the effective logarithmic mass obtained from two-point correlators for three baryons—the nucleon, $\Omega$, and $\Omega_c$—on two representative ensembles: the physical point ensemble C48P14 and the finest lattice spacing ensemble H48P32. The results are shown using the Coulomb wall source (red crosses) and, where available, the distillation method (black circles) from CLQCD collaborators.

For both the nucleon and $\Omega$ baryons, we observe that the effective masses from the Coulomb wall source and distillation method are in good agreement once the plateau is reached, even though the initial trends with respect to $t/a$ differ; specifically, the Coulomb wall source tends to increase with $t$, while the distillation results exhibit a decrease before stabilizing. This consistency at the plateau confirms the reliability of both approaches. Even though no comparative distillation data is available for the $\Omega_c$ baryon, the statistical uncertainty here is small enough to ensure the reliability of the fit. The green (blue) bands in each plot indicate the fit uncertainties for the extracted effective masses which consistent with the original data points well.

\subsection{Joint fit of the baryon mass}\label{sec:extrapolation}
We fit the nucleon mass using the the SU($4|2$) partially quenched heavy baryon chiral perturbative theory ($\chi$PT) ansatz~\cite{Tiburzi:2005is,Yang:2018nqn},
\begin{align}
m_H\left(m_\pi^{\rm val}, m_\pi^{\mathrm{sea}}, m_{\eta_s}^{\mathrm{sea}}, a, 1/L\right) =& m_H^{\rm phys}
+ \sum_{{\rm tag=val/pq}, j=2,3}C^{\rm tag}_{j} [\left(m_\pi^{\rm tag}\right)^j-\left(m_\pi^{\rm phys}\right)^j]
+ C_s \left[(m_{\eta_s}^{\mathrm{sea}})^2 - (m_{\eta_s}^{\mathrm{phys}})^2\right] \nonumber\\
&\quad+ C_L \frac{\left(m_\pi^{\rm val}\right)^2}{L} e^{-m_\pi^{\rm val} L} 
 + \sum_{i}C^{2i}_{a} a^{2i}\;,
\label{eq:baryon}
\end{align}
where $m_\pi^{\rm val}$ and $m_\pi^{\rm sea}$ are the valence and sea pion masses respectively, $m^{\mathrm{pq}} = \sqrt{\left(m_\pi^{\rm{ val}}\right)^2 + \left(m_\pi^{\mathrm{sea}}\right)^2}$ is the partially quenched pion mass, $m_{\pi}^{\rm phys}=134.98$ MeV is the physical $\pi^0$ mass, and $m_{\eta_s}^{\mathrm{phys}}=689.89(49)$ MeV from Ref.~\cite{Borsanyi:2020mff} is the pure QCD $\eta_s$ mass which corresponds to the physical strange quark mass.

\begin{figure}[thb]
    \includegraphics[width=0.45\textwidth]{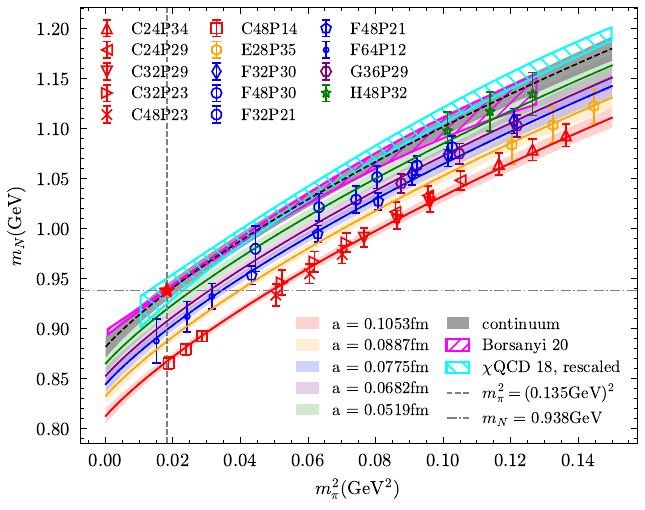}
    \caption{The nucleon mass on 14 ensembles at 5 lattice spacings (colored data points and bands), and also that in the continuum (gray band). The mismatch effect from the unequal valence and sea pion masses has been corrected based on the joint fit. The continuum extrapolated results using the overlap fermion~\cite{Yang:2018nqn} (rescaled to match the physical nucleon mass) and staggered fermion~\cite{Borsanyi:2020bpd} are also illustrated for comparison.}
    \label{fig:proton_masses}
\end{figure}

In Fig.~\ref{fig:proton_masses}, we present the nucleon mass at five different lattice spacings as colored data points, with the mismatch effects between $m_\pi^{\rm val}$ and $m_\pi^{\rm sea}$ are corrected using the joint fit which gives $\chi^2$/d.o.f.$=1.2$. Redefinition of parameters $C^{\rm val}_3=\frac{\left(g_A^2 - 4 g_A g_1 - 5 g_1^2\right) \pi}{3 \left(4 \pi f_\pi\right)^2}$ and $C^{\rm pq}_3=\frac{\left(8 g_A^2 + 4 g_A g_1 + 5 g_1^2\right) \pi}{3 \left(4 \pi f_\pi\right)^2}$ are used to be consistent with the previous studies~\cite{Tiburzi:2005is,Yang:2018nqn}. The colored bands corresponds to the fit prediction at respective lattice spacings, and they agree with the data points well. Based on the linear $a^2$ extrapolation, our continuum extrapolated result as function of pion mass is shown as gray band, which perfectly agree with the that from  using the overlap fermion~\cite{Yang:2018nqn} and staggered fermion~\cite{Borsanyi:2020bpd}. The physical nucleon mass is shown as the red star. Note that the band from Ref.~\cite{Yang:2018nqn} is rescaled by a factor 0.978(16) from the original data to match the physical point.

For the light baryons containing a strange quark, we calculate the baryon masses using three different valance strange quark masses for each ensemble. We then interpolate the valance strange quark mass to the value corresponding to the artificial $\eta_s$ (unmixed $\bar{s}s$) mass $m_{\eta_s}^{\mathrm{phys}} = 689.89(49)\,\mathrm{MeV}$~\cite{Borsanyi:2020mff}, which is determined by using the physical strange quark mass and is consistent with the value of $687.4(2.2)\,\mathrm{MeV}$~\cite{CLQCD:2023sdb} obtained from CLQCD ensembles, albeit with a larger uncertainty. We will revisit this strategy in the next section. The $m_{\pi}^3$ terms are omitted because $\chi$PT is not applicable here, and their coefficients show no significant signal in any baryon containing strange or charm valence quarks.  

The cases of charmed baryons are also similar. Following the the GRS renormalization scheme $m^{\overline{\textrm{MS}}}_{q, \rm QCD+QED}(2\mathrm{GeV})=m^{\overline{\textrm{MS}}}_{q, \rm QCD}(2\mathrm{GeV})$, we use the pure QCD value $m_{D_s}=1966.7(1.5)\,\mathrm{MeV}$~\cite{DiCarlo:2019thl} to determine the interpolated valence charm quark mass for each ensemble. The resulting masses of open and closed charmed mesons, as well as their decay constants, agree well with experimental values and lattice averages \cite{CLQCD:2024yyn}.

The interpolation of valence strange and charm quark masses significantly suppresses the discretization errors from the heavy quark, and also allows us to use a similar ansatz as that used for nucleons to fit the baryon mass for different flavors and angular momentum. For the charmed baryon, we retain the $a^4$ term in the extrapolation, as the linear $a^2$ extrapolation does not describe the data well, resulting in poor $\chi^2$. We also take $m_\pi^{\rm val}=m_\pi^{\rm sea}$ and drop the $\left(m_\pi^{\rm tag}\right)^{2,3}$ terms for baryon without valence light quark.

\begin{figure}[thb]    
\includegraphics[width=0.45\textwidth]{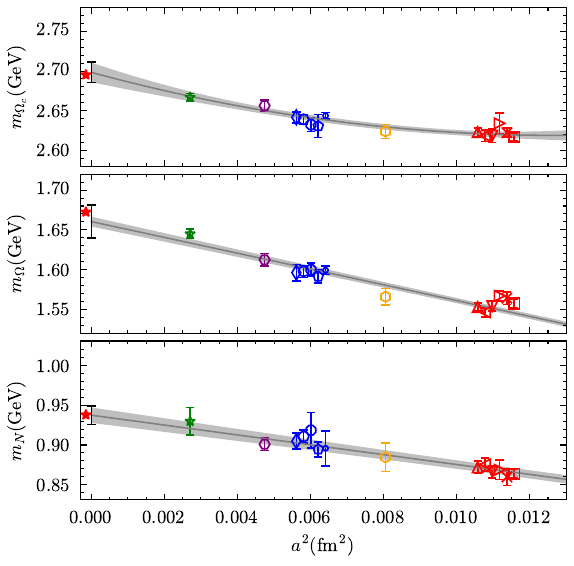}
    \caption{Fitting results for nucleon, $\Omega$, and $\Omega_c$ masses as functions of $a^2$. The other non-physical effects for the data points are fixed using the global fit. For lattice spacings with more than one ensemble, data points are horizontally displaced for clarity. }
    \label{fig:discretization_effects}
\end{figure}
In Fig.~\ref{fig:discretization_effects}, we show the lattice spacing dependence of the nucleon, $\Omega$ and $\Omega_c$ masses, with the light and strange quark masses corrected to their physical value based on the joint fit on 14 ensembles. The gray bands indicate the fit uncertainties, and the red stars correspond to the experimental values. Obviously the linear $a^2$ dependence describe the data of nucleon and $\Omega$ well, while the $a^4$ term is essential fo the $\Omega_c$.  For a comprehensive view, we present the results for all baryons in Fig.~\ref{fig:all_discretization_effects}, whrere we apply an $a^2$ extrapolation for light baryons, while for charmed baryons, an $a^4$ extrapolation is used to capture higher-order discretization effects.

\begin{figure}[h]
\includegraphics[width=\textwidth]{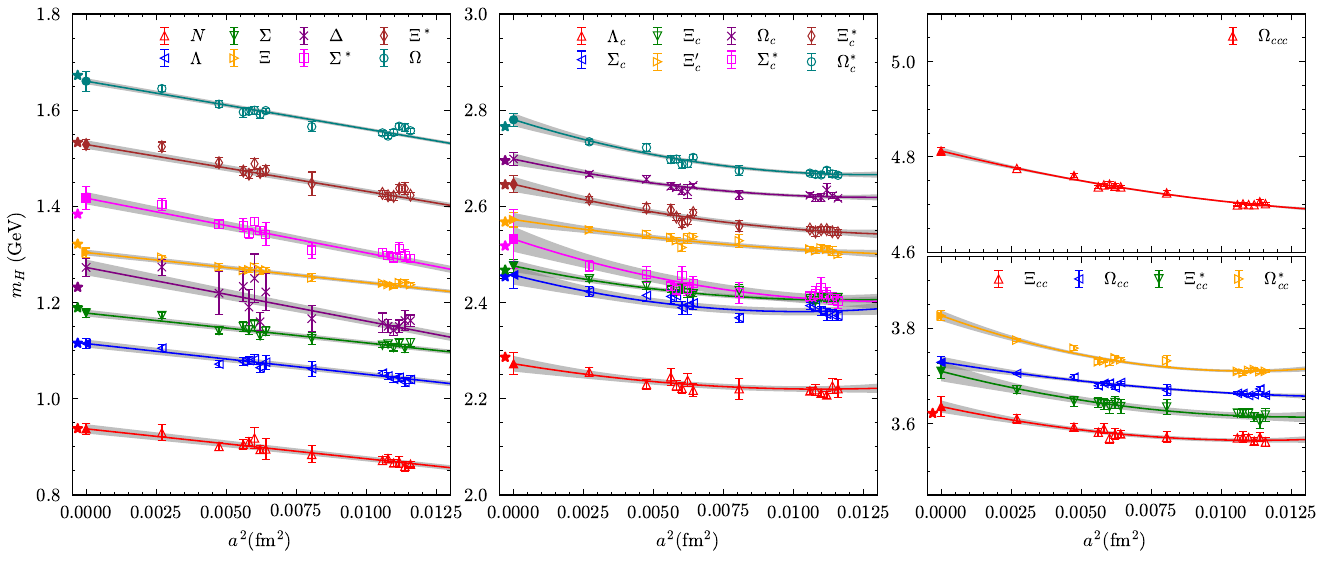}
\caption{Fitting results for the masses of all baryons, showing their dependence on lattice spacing $a^2$. For light baryons, an $a^2$ extrapolation is applied, while for charmed baryons, an $a^4$ extrapolation is used where necessary. The gray bands represent fit uncertainties, and colored stars indicate experimental values when available. All data points have been corrected for non-physical effects using the global fit, demonstrating consistent treatment of discretization effects across the baryon spectrum. For lattice spacings with more than one ensemble, data points are horizontally displaced for clarity.}
    \label{fig:all_discretization_effects}
\end{figure}

\subsection{$\sigma_{q H}$ extraction}

The Feynman-Hellman theorem establishes the relation $\langle \bar{q}q\rangle_H=\frac{\partial m_H}{\partial m_q}$,
which can be further decomposed into contributions from valence and sea quarks:
\begin{align}
    \langle \bar{q}q\rangle_H=\langle \bar{q}q\rangle_H^{\rm val}+\langle \bar{q}q\rangle_H^{\rm sea},\ \langle \bar{q}q\rangle_H^{\rm val}=\frac{\partial m_H(m^{\rm val}_q,m^{\rm sea}_q)}{\partial m^{\rm val}_q},\ \langle \bar{q}q\rangle_H^{\rm sea}=\frac{\partial m_H(m^{\rm val}_q,m^{\rm sea}_q)}{\partial m^{\rm sea}_q}.
\end{align}
Here, \(m_q^{\rm val/sea}\) denote the valence/sea quark masses, and \(m_H(m^{\rm val}_q, m^{\rm sea}_q)\) is derived from joint fits of lattice data across ensembles with varying \(m_q^{\rm val}\) and \(m_q^{\rm sea}\). The valence term \(\langle \bar{q}q\rangle_H^{\rm val}\) can first be extracted on individual ensembles and then extrapolated to the physical quark masses, infinite volume, and continuum limit through a global fit.  

Since the valence strange and charm quark masses (\(m_{s,c}^{\rm val}\)) are tuned to reproduce the pure-QCD masses \(m_{\eta_s}\) and \(m_{D_s}\) on each ensemble (where sea quark masses \(m_{l,s}^{\rm sea}\) may deviate from physical values), mismatches \(\delta m^s_{l,s} \equiv m_{l,s}^{\mathrm{sea}} - m_{l,s}^{\rm phys}\) induce corrections to the valence masses: \(\mathbf{\delta m_{s,c}^{\mathrm{val}}} \equiv m_{s,c}^{\mathrm{val}} - m_{s,c}^{\rm phys}\) with the following constraints:  
\begin{align}
    0=\delta m_{\eta_s}&= \delta m^s_l \langle\bar{l}l\rangle^{\mathrm{sea}}_{\eta_s}+\mathbf{\delta m_s^{\mathrm{val}} }\langle\bar{s}s\rangle^{\mathrm{val}}_{\eta_s}+\delta m_s^{\mathrm{sea}}\langle\bar{s}s\rangle^{\mathrm{sea}}_{\eta_s},\nonumber\\
    0=\delta m_{D_s}&= \delta m^s_l \langle\bar{l}l\rangle^{\mathrm{sea}}_{D_s}+\mathbf{\delta m_s^{\mathrm{val}}}\langle\bar{s}s\rangle^{\mathrm{val}}_{D_s}+\delta m_s^{\mathrm{sea}}\langle\bar{s}s\rangle^{\mathrm{sea}}_{D_s}+\mathbf{\delta m_c^{\mathrm{val}}} \langle\bar{c}c\rangle^{\mathrm{val}}_{D_s},\label{eqn:constraints}
\end{align}
where all the condensates correspond to those at physical quark masses. The resulting correction to any hadron mass \(m_H\) is: 
\begin{align}
    \delta m_{H}&= \delta m^{\mathrm{sea}}_l \langle\bar{l}l\rangle^{\mathrm{sea}}_{H}+\mathbf{\delta m_s^{\mathrm{val}}}\langle\bar{s}s\rangle^{\mathrm{val}}_{H}+\delta m_s^{\mathrm{sea}}\langle\bar{s}s\rangle^{\mathrm{sea}}_{H}+\mathbf{\delta m_c^{\mathrm{val}}} \langle\bar{c}c\rangle^{\mathrm{val}}_{H}\;\nonumber\\
    &=\delta m^{\mathrm{sea}}_l \left( \langle\bar{l}l\rangle^{\mathrm{sea}}_{H} + \frac{\partial \mathbf{m_s^{\mathrm{val}}}}{\partial m^{\mathrm{sea}}_l} \langle\bar{s}s\rangle^{\mathrm{val}}_{H} + \frac{\partial \mathbf{m_c^{\mathrm{val}}}}{\partial m^{\mathrm{sea}}_l} \langle\bar{c}c\rangle^{\mathrm{val}}_{H} \right)  + \delta m_s^{\mathrm{sea}} \left( \langle\bar{s}s\rangle^{\mathrm{sea}}_{H} + \frac{\partial \mathbf{m_s^{\mathrm{val}}}}{\partial m^{\mathrm{sea}}_s} \langle\bar{s}s\rangle^{\mathrm{val}}_{H} + \frac{\partial \mathbf{m_c^{\mathrm{val}}}}{\partial m^{\mathrm{sea}}_s} \langle\bar{c}c\rangle^{\mathrm{val}}_{H} \right).
\end{align}
Thus, when extracting \(\langle\bar{l}l\rangle^{\mathrm{sea}}_{H}\) from the light quark mass dependence of \(m_H\), contamination from \(\mathbf{\delta m_{s,c}^{\mathrm{val}}}\) must be subtracted.  

After this subtraction, the sigma terms used in this work are defined as:  
\begin{align}
    \sigma_{\pi H}&=\sigma^{\rm val}_{\pi H}+\sigma^{\rm sea}_{\pi H},\nonumber\\
    \sigma^{\rm val,H}_{\pi H}&\equiv (m_{\pi}^{\rm phys})^2\frac{\partial m_H(m^{\rm val}_{\pi},m^{\rm phys}_{\pi},m^{\rm phys}_{\eta_s},0,0)}{\partial (m^{\rm val}_{\pi})^2}\bigg|_{m^{\rm val}_{\pi}=m^{\rm phys}_{\pi}},\\
    \sigma^{\rm sea}_{\pi H}&\equiv (m_{\pi}^{\rm phys})^2\big[\frac{\partial m_H(m^{\rm phys}_{\pi},m^{\rm sea}_{\pi},m^{\rm phys}_{\eta_s},0,0)}{\partial (m^{\rm sea}_{\pi})^2}\bigg|_{m^{\rm sea}_{\pi}}-d_l^{s}\frac{\sigma^{\rm val}_{s,H}}{m^{\rm phys}_s}-d_l^{c}\frac{\sigma^{\rm val}_{c,H}}{m^{\rm phys}_c}\big],\label{eq:sea_pi}\\
    \sigma_{sH}&=\sigma^{\rm val}_{sH}+\sigma^{\rm sea}_{sH},\nonumber\\
    \sigma^{\rm val}_{sH}&\equiv \sigma^{\rm val}_{sH}(m^{\rm sea}_{\pi},m^{\rm phys}_{\eta_s},0,0),\label{eq:sigma_s_v}\\
    \sigma^{\rm sea}_{sH}&\equiv (m_{\eta_s}^{\rm phys})^2\big[\frac{\partial m_H(m^{\rm phys}_{\pi},m^{\rm phys}_{\pi},m^{\rm sea}_{\eta_s},0,0)}{\partial (m^{\rm sea}_{\eta_s})^2}\bigg|_{m^{\rm sea}_{\eta_s}}-d_s^{s}\frac{\sigma^{\rm val}_{s,H}}{m^{\rm phys}_s}-d_s^{c}\frac{\sigma^{\rm val}_{c,H}}{m^{\rm phys}_c}\big],\label{eq:sea_str}\\
    \sigma^{\rm val}_{cH}&\equiv \sigma^{\rm val}_{cH}(m^{\rm sea}_{\pi},m^{\rm phys}_{\eta_s},0,0),\label{eq:sigma_c_v}
\end{align}
where $m_H$ is defined in Eq.~(\ref{eq:baryon}) with tuned valence strange and charm quark masses.

\begin{figure}[thb]
\includegraphics[width=0.45\textwidth]{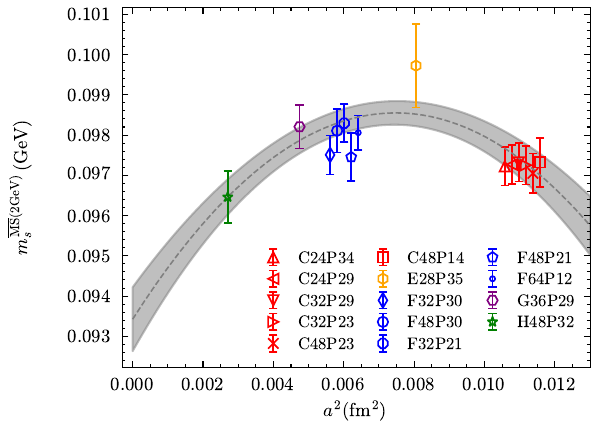}
    \caption{Lattice spacing dependence of the renormalized strange quark mass at $\overline{\mathrm{MS}}$(2~GeV). The values at finite lattice spacings are corrected to the physical pion and $\eta_s$ masses based on the joint fit. For lattice spacings with more than one ensemble, data points are horizontally displaced for clarity.}
    \label{fig:strange}
\end{figure}

The coefficients \(d^{s(c)}_{l(s)}\) arise from joint fits of the renormalized strange and charm quark masses:  
\begin{align}
    m^{\rm val,R}_{s(c)} = m^{\rm phys,R}_{s,c} + d_l^{s(c)} \left( (m_{\pi}^\mathrm{sea})^2 - \left( m_{\pi}^{\rm phys} \right)^2 \right) + d^{s(c)}_s \left( (m_{\eta_s}^{\mathrm{sea}})^2 - \left( m_{\eta_s}^{\rm phys} \right)^2 \right)+\sum_{i=1}^2d^{s(c)}_{a,i} a^{2i} +  d^{s(c)}_L e^{-m_{\pi}L}.\label{eq:mass_dep}
\end{align}
For the charm quark, we obtain \(d_l^c=-0.161(11)\,\rm{(GeV)}^{-1}\), \(d_s^c=-0.388(44)\,\rm{(GeV)}^{-1}\), and a physical renormalized mass \(m^{\rm phys,R}_c = 1.0989(89)\)\,GeV in the \(\overline{\mathrm{MS}}\) scheme at 2\,GeV. A similar fit for the strange quark yields \(d_l^s = -0.0267(30)\,\rm{(GeV)}^{-1}\), \(d_s^s=-0.0090(10)\,\rm{(GeV)}^{-1}\), and \(m_s^{\rm phys,R}=0.09345(79)\)\,GeV in the \(\overline{\mathrm{MS}}\) scheme at 2\,GeV. As shown in Fig.~\ref{fig:strange}, the \(a^4\) discretization correction to \(m_s^{\rm phys,R}\) based on $m_{\eta_s}$ is also significant, while the continuum extrapolated \(m_s^{\rm phys,R}\) is still consistent with our previous estimate 0.0978(21)(40) MeV in Ref.~\cite{CLQCD:2023sdb}.

The valence strange (charm) sigma term $\sigma^{\rm val}_{s(c)H}$ defined in Eqs.~(\ref{eq:sigma_s_v}) and (\ref{eq:sigma_c_v}) are extracted from the joint fit of the sigma term at each ensemble 
using the Feynman-Hellman theorem with the renormalized valence quark mass $m_{s(c)}^{\rm val, R}$, 
\begin{align}
\sigma^{\rm val}_{s(c)H}(m^{\rm sea}_{\pi},m^{\rm sea}_{\eta_s},a,1/L)&\equiv m_{s(c)}^{\rm val, R}(m^{\rm sea}_{\pi},m^{\rm sea}_{\eta_s},a,1/L) \frac{\partial m_H(m^{\rm sea}_{\pi},m^{\rm sea}_{\eta_s},a,1/L)}{\partial m_{s(c)}^{\rm val, R}(m^{\rm sea}_{\pi},m^{\rm sea}_{\eta_s},a,1/L)}\nonumber\\
&=\sigma^{\rm val}_{s(c) H}(m_{\pi}^{\rm phys},m_{\eta_s}^{\rm phys},0,0) +C^{\sigma_{s(c)H}}_l ((m_{\pi}^\mathrm{sea})^2-(m_{\pi}^{\rm phys})^2)+C^{\sigma_{s(c)H}}_s ((m^{\rm sea}_{\eta_s})^2-(m_{\eta_s}^{\rm phys})^2)\nonumber\\
    &\quad \quad \quad +\sum_iC^{\sigma_{s(c)H}}_{a,i} a^{2i}+  C^{\sigma_{s(c)H}}_L e^{-m_{\pi}L} ,\label{eq:sigma_dep} 
\end{align}
Discretization errors for \(\sigma^{\rm val}_{sH}\) are modeled with \(a^2\) terms, while both \(a^2\) and \(a^4\) terms are needed for \(\sigma^{\rm val}_{cH}\). 

In above expressions, the \(m_{\pi}^2\) dependence approximates the light quark mass (\(m_l\)) dependence, as higher-order chiral corrections (\(\sim 2\%\))~\cite{CLQCD:2023sdb} are negligible compared to statistical uncertainties. Similarly, \(m_{\eta_s}^2/m_s \approx 4.9(1)\)~\cite{CLQCD:2023sdb} near the physical \(m_s\) justifies using \(m_{\eta_s}^2\) to model \(m_s\) dependence.

\begin{figure}[thb]
\includegraphics[width=0.98\textwidth]{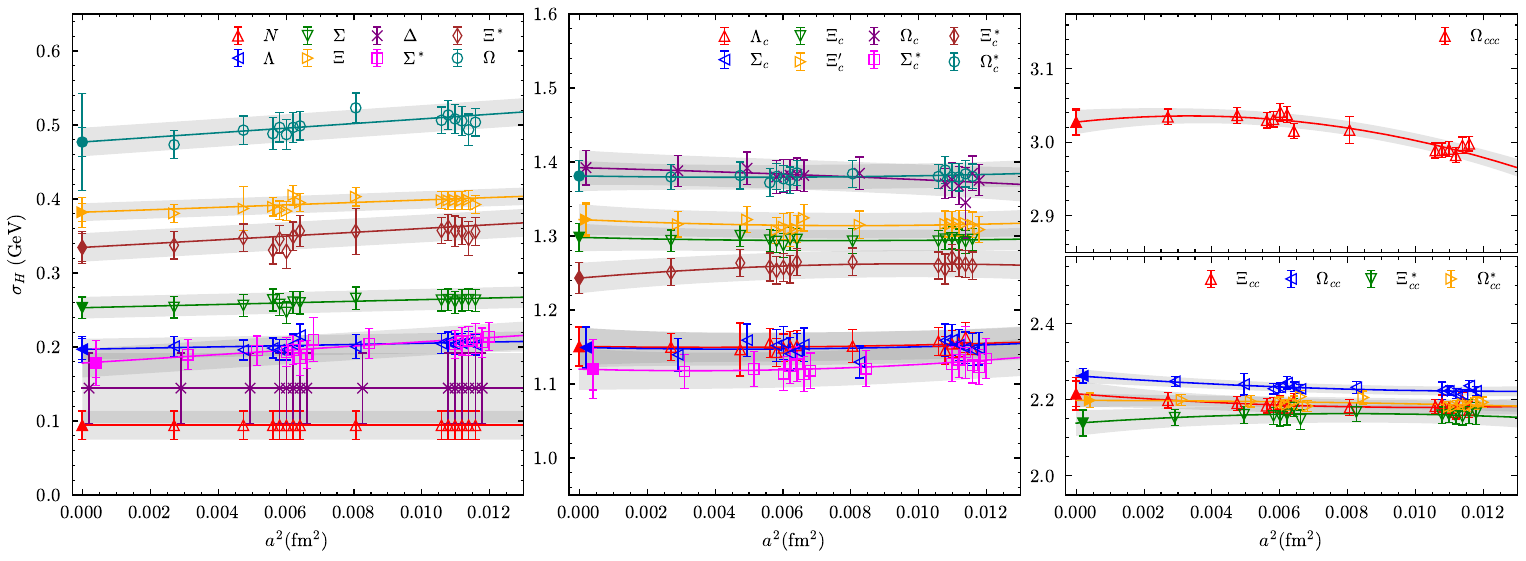}
    \caption{Similar to Fig.~\ref{fig:all_discretization_effects} but for $\sigma_H$. The plots indicate that the discretization effects on $\sigma_H$ are generally smaller than those observed for the baryon masses, though the associated uncertainties are comparatively larger. Data points are horizontally displaced for clarity.}
    \label{fig:Hm_all_particles_appendix}
\end{figure}

In Fig.~\ref{fig:Hm_all_particles_appendix}, we present the plots of $\sigma_H$ for all baryons, similar to Fig.~\ref{fig:all_discretization_effects} for the baryon masses. As indicated in the figure, the discretization effects on $\sigma_H$ are generally smaller than those on the total baryon mass, highlighting the stability of $\sigma_H$ under different lattice spacings.

\begin{figure*}[thb]
    \includegraphics[width=0.98\textwidth]{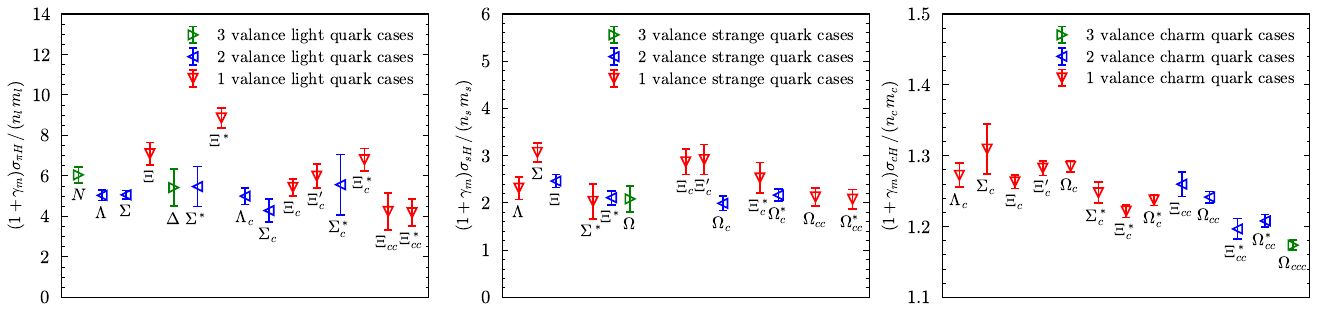}
    \caption{Ratio $(1+\gamma_m)\sigma_{q,H}/(n_qm_q)$ for the light (left), strange (middle) and charm (right) quark in different hadron $H$, at $\overline{\textrm{MS}}$ 2~GeV.}
    \label{fig:ratio_3in1}
\end{figure*}

In Fig.~\ref{fig:ratio_3in1}, we present the ratio \( R_{q,H} \equiv (1+\gamma_m)\sigma_{q,H}/(n_q m_q) \) for various quark flavors \( q \) in different hadrons \( H \), where \( n_q \) denotes the number of valence quarks of flavor \( q \). The results are given in the \(\overline{\textrm{MS}}\) scheme at 2 GeV. Although \( R_{q,H} \) itself depends on the renormalization scale and scheme, the \textbf{relative} ratio \( R_{q,H} \) across different hadrons and quark flavors is scale- and scheme-independent. If we normalize all \( R_{q,H} \) values by \( R_{c\Omega_{ccc}} \), we find that \( R_{sH}/R_{c\Omega_{ccc}} \approx 2 \) for all baryons, while \( R_{\pi H}/R_{c\Omega_{ccc}} \) is even larger.

\begin{figure*}[thb]
    \includegraphics[width=0.98\textwidth]{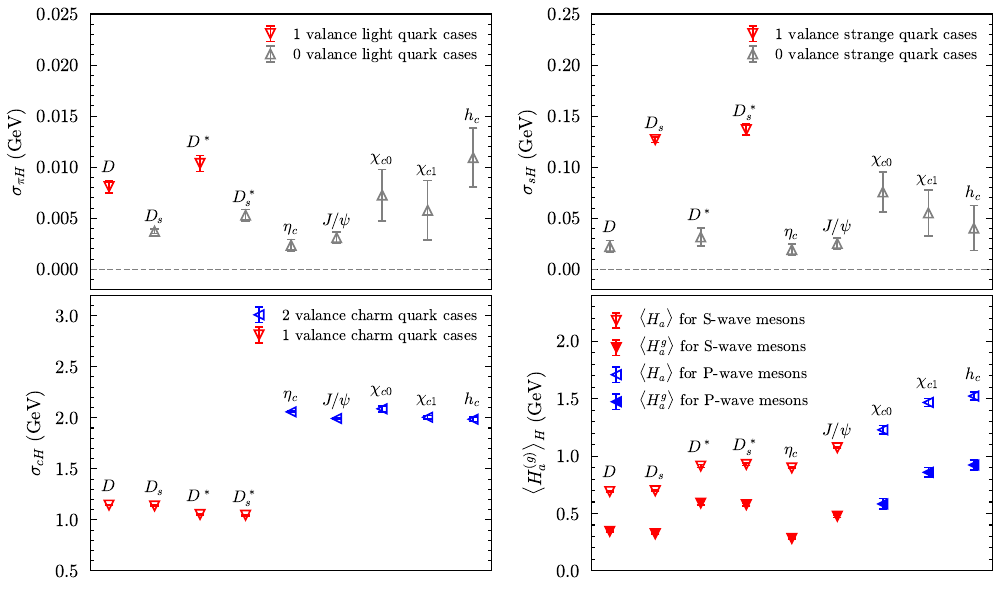}
    \caption{Results for quark sigma terms and trace anomaly contributions to charmed meson masses. The upper-left, upper-right, and lower-left panels show the $\sigma$ terms for light, strange, and charm quarks, respectively, with colors indicating the number of valence quark (and anti-quark): gray (0), red (1), blue (2). The lower-right panel presents the trace anomaly $\langle H_a \rangle_H$ and the gluon trace anomaly $\langle H^g_a \rangle_H$, with red triangles for S-wave meson and blue triangles for P-wave meson. }
    \label{fig:Hm_4in1_meson}
\end{figure*}

Based on the similar procedure, we can also obtain the sigma term and trace anomaly for the charmed meson discussed in Ref.~\cite{CLQCD:2024yyn}. The comparison of $\sigma_{\pi H}$ for hadrons $H$ with and without valence light quarks reveals that valence and sea contributions are comparable, as shown in Fig.~\ref{fig:Hm_4in1_meson}. We also observe that the gluon trace anomaly of pseudoscalar mesons is generally smaller than that of baryons after subtracting the $\gamma_m$ contribution at $\overline{\textrm{MS}}$ 2~GeV. Simultaneously, the trace anomalies of vector and P-wave charmed mesons appear somewhat larger, though none exceed 1~GeV like the $3/2^+$ baryon case. Further investigation of other excited charmed hadrons would be valuable for understanding the role of trace anomaly in their mass generation.

\subsection{Error budget of our predictions}

In table \ref{tab:results}, we summarized the final numerical results of the baryon masses, sigma terms and also trace anomaly contributions. 

The statistical uncertainty is categorized by its sources: the statistical fluctuations in the dimensionless pion and $\eta_s$ masses in the sea quark sector, the lattice spacing determination using gradient flow (excluding systematic effects from the scale parameter v taken from literature ~\cite{flavourlatticeaveraginggroupflag:2024oxs}), and those from the dimensionless two-point correlation functions.

For systematic uncertainties related to the scale parameter $w_0$= 0.1736(9) fm~\cite{flavourlatticeaveraginggroupflag:2024oxs}), we perform a complete reanalysis by varying the central w0 value by 1$\sigma$ and take the resulting difference in central values as the systematic uncertainty from lattice spacing. Additional systematic uncertainties are considered from:

1) The experimental $D_s$ mass measurement;

2) Choice of fit ansatz (additive vs. multiplicative discretization errors);

3) Sea quark dependence parameters $d_{1,2}^{s(c)}$ in Eqs.~(\ref{eq:sea_pi}) and (\ref{eq:sea_str}), which propagate to uncertainties of the sea quark mass contribution of sigma terms, and then also that of the gluon trace anomaly due to the sum rule.

The primary source of uncertainty is statistical, arising from the original two-point function, while scale setting can also contribute significantly to systematic uncertainty, particularly for charmed hadrons.

\begin{longtable}{l l c c c c c c c c c c c}
    \caption{Error budget for baryon masses, sigma terms, and trace anomalies (all in GeV).} \label{tab:results}\\
    &  &$m_H$& $\sigma^{\rm val}_{\pi H}$& $\sigma^{\rm sea}_{\pi H}$& $\sigma_{\pi H}$ &  $\sigma^{\rm val}_{s H}$&  $\sigma^{\rm sea}_{s H}$&  $\sigma_{s H}$ & $\sigma^{\rm val}_{c H}$& $\sigma_{ H}$ & $\langle H_a \rangle_H$ & $\langle H_a^g \rangle_H$\\
    \endfirsthead

    &  &$m_H$& $\sigma^{\rm val}_{\pi H}$& $\sigma^{\rm sea}_{\pi H}$& $\sigma_{\pi H}$ &  $\sigma^{\rm val}_{s H}$&  $\sigma^{\rm sea}_{s H}$&  $\sigma_{s H}$ & $\sigma^{\rm val}_{c H}$& $\sigma_{ H}$ & $\langle H_a \rangle_H$ & $\langle H_a^g \rangle_H$\\
    \hline
    \endhead

    \multicolumn{13}{r}{\textit{Continued on next page}} \\
    \hline
    \endfoot

    \endlastfoot
\hline
  \multirow{1}{*}{$N$}  & Central value & 0.9376 & 0.0291 & 0.0184 & 0.0475 & 0 & 0.0468 & 0.0468 & - & 0.0942 & 0.8434 & 0.8156 \\
    \cline{2-13}
     & \textbf{Statistical (total)} & 0.0098 & 0.0017 & 0.0019 & 0.0025 & 0 & 0.0176 & 0.0176 & - & 0.0194 & 0.0217 & 0.0270 \\
    & Sea $\pi$ mass & 0.0006 & \textless 0.0001 & 0.0001 & \textless 0.0001 & 0 & 0.0004 & 0.0004 & - & 0.0004 & 0.0007 & 0.0008 \\
    & Sea $\eta_s$ mass & 0.0025 & \textless 0.0001 & \textless 0.0001 & \textless 0.0001 & 0 & 0.0008 & 0.0008 & - & 0.0007 & 0.0026 & 0.0026 \\
    & Lattice spacing & 0.0015 & 0.0001 & \textless 0.0001 & 0.0001 & 0 & 0.0007 & 0.0007 & - & 0.0009 & 0.0012 & 0.0013 \\
    & 2-point function & $\mathbf{0.0093}$ & $\mathbf{0.0017}$ & $\mathbf{0.0019}$ & $\mathbf{0.0025}$ & 0 & $\mathbf{0.0176}$ & $\mathbf{0.0176}$ & - & $\mathbf{0.0194}$ & $\mathbf{0.0215}$ & $\mathbf{0.0268}$ \\
    \cline{2-13}
    & \textbf{Systematic (total)} & 0.0065 & 0.0009 & 0.0008 & 0.0018 & 0 & 0.0007 & 0.0007 & - & 0.0022 & 0.0087 & 0.0093 \\
    & Scale setting & 0.0042 & 0.0002 & \textless 0.0001 & 0.0002 & 0 & $\mathbf{0.0006}$ & $\mathbf{0.0006}$ & - & 0.0009 & 0.0051 & 0.0054 \\
\hline
\end{longtable}

\end{widetext}

